\documentclass{jltp}

\usepackage{graphicx} 

\title{Scanning Tunneling Spectroscopic Studies of the Pairing State
of Cuprate Superconductors}

\author{N.-C. Yeh, C.-T. Chen, R. P.
Vasquez$^*$, C. U. Jung$^\dag$, S.-I. Lee$^\dag$,\\
K. Yoshida$^\ddag$ and S. Tajima$^\ddag$}

\address{Department of Physics, California Institute of
Technology, Pasadena,\\CA 91125, USA\\ $^*$Jet Propulsion
Laboratory, California Institute of Technology,
Pasadena,\\CA 91109, USA\\
$^\dag$Department of Physics, Pohang University of Science and
Technology,\\Pohang 790-784, Korea\\
$^\ddag$Superconductivity Research Laboratory, International
Superconductivity\\ Technology Center, Sinonome, Koto-ku, Tokyo,
135 Japan}

\runninghead{\bf N.-C.~Yeh \boldmath $et\ al.$}{Scanning Tunneling Spectroscopic Studies of Cuprate Superconductors}

\begin{document}

\maketitle

\begin{abstract}
Quasiparticle tunneling spectra of both hole-doped (p-type) and
electron-doped (n-type) cuprates are studied using a
low-temperature scanning tunneling microscope. The results reveal
that neither the pairing symmetry nor the pseudogap phenomenon is
universal among all cuprates, and that the response of n-type
cuprates to quantum impurities is drastically different from that
of the p-type cuprates. The only ubiquitous features among all
cuprates appear to be the strong electronic correlation and the
nearest-neighbor antiferromagnetic Cu$^{2+}$-Cu$^{2+}$ coupling in
the CuO$_2$ planes.

\end{abstract}

\section{INTRODUCTION}
To date there has been no consensus for the mechanism of cuprate
superconductivity. The diverging views are primarily due to
complications incurred by competing orders in these strongly
correlated doped Mott insulators.~\cite{Sachdev00,Vojta00,Lee02}
The competing orders can result in a variety of phases in the
ground state, depending on whether the cuprate is hole doped
(p-type) or electron doped (n-type), and also on the carrier
doping level, the electronic coupling strength between neighboring
CuO$_2$ planes, and the degree of
disorder.\cite{Yeh01a,Yeh01b,Yeh02,Chen02} In order to sort
through the complications, it is necessary to identify universal
characteristics among all cuprate superconductors.

Among the noteworthy phenomena associated with the cuprates,
$d_{x^2-y^2}$-wave pairing symmetry,\cite{vanHarlingen95,Tsuei00a}
spin fluctuations,\cite{Pines97a} pseudogap
phenomenon,\cite{Timusk99} strong phase fluctuations\cite{Emery95}
and stripe phases\cite{Siebold98} have been considered as
important to the underlying mechanism. In particular, the
pseudogap phenomenon is widely regarded as a precursor to
superconductivity, and its doping dependence together with the
non-Fermi liquid behavior in the pseudogap regime have led to the
hypothesis for a quantum critical point (QCP) near the optimal
doping level\cite{Varma97} as well as other models such as
preformed Cooper pairs,\cite{Emery97} Bose-Einstein condensation
at $T_c$,\cite{Uemura89} $d_{x^2-y^2}$-density wave phase with
orbital currents in the pseudogap regime,\cite{Chakravarty01} and
spin gap scenario.~\cite{Lee97} However, these models have been
developed around experimental findings in the p-type cuprates.
Given that the cuprates cannot be properly described by a simple
one-band Hubbard model and therefore lack apparent particle-hole
symmetry, it is not obvious whether all phenomena in the p-type
cuprates can be generalized to the n-type cuprates.

In this work, we compare the quasiparticle tunneling spectra of a
variety of p-type and n-type cuprate superconductors, with special
emphasis on the doping dependence and the effects of quantum
impurities in the CuO$_2$ planes. The physical implications of our
investigation are discussed.

\section{PAIRING STATE OF THE P-TYPE CUPRATES}
The p-type cuprates studied in this work include the under- and
optimally doped $\rm YBa_2Cu_3O_{7-\delta}$ (YBCO), overdoped YBCO
with Ca-substitution, an optimally doped YBCO with 0.26\% Zn and
0.4\% Mg substitutions for Cu in the CuO$_2$
planes,\cite{Yeh01a,Yeh01b,Yeh02} and the under- and optimally
doped $\rm La_{2-x}Sr_xCuO_{4-\delta}$ (LSCO) system.\cite{Wei00}
The spectra are taken using a low-temperature scanning tunneling
microscope (STM), with the average quasiparticle momentum along
different crystalline axes. The surface preparation with chemical
etching method and surface characterizations are described
elsewhere.\cite{Yeh01a,Vasquez88,Vasquez01}

\subsection{Doping-Dependent Pairing Symmetry and Pairing Potential}
The pairing symmetry and momentum ($k$)-dependent pairing
potential $\Delta _k$ can be derived self-consistently by applying
the generalized BTK analysis to the directional tunneling spectra
taken on each sample.\cite{Yeh01a,Yeh01b,Wei98} We find that the
pairing symmetry of p-type cuprates is dependent on the doping
level and the crystalline structure. For cuprates with
orthorhombic structure such as the YBCO system, the pairing
symmetry is predominantly $d_{x^2-y^2}$-wave in the underdoped
regime and ($d_{x^2-y^2}+s$)-wave with a significant $s$-wave
component increasing with doping in the overdoped
limit.\cite{Yeh01a,Yeh01b,Yeh02} On the other hand, for tetragonal
samples such as the $\rm Bi_2Sr_2CaCu_2O_x$ (BSCCO) and LSCO, the
pairing symmetry is consistent with pure $d_{x^2-y^2}$-wave for
all doping levels. Moreover, no discernible complex order
parameter could be detected in the p-type cuprates based on our
STS studies.\cite{Yeh01a,Yeh01b,Yeh02} Therefore no obvious QCP
with a universal broken symmetry can be identified in the p-type
cuprates based on our tunneling studies. In addition, the
$d_{x^2-y^2}$-component of the superconducting gap ($\Delta _d$)
in the YBCO system is only weakly dependent on the doping level
for the optimal and underdoped samples,\cite{Yeh01a,Yeh01b,Yeh02}
and is reduced significantly in the overdoped regime, as shown in
Fig.~\ref{fig1}(a). These gap values do not persist above
$T_c$,\cite{Maggio00} and are therefore consistent with the
pairing potential. Our findings are in contrast to the
point-contact spectra of BSCCO that exhibit a rapidly increasing
averaged energy gap with decreasing doping,\cite{Miyakawa99} and
the large averaged gap of the underdoped BSCCO appears to merge
with the pseudogap. However, recent tunneling experiments on BSCCO
mesas suggest that the spectral pseudogap does not have the same
physical origin as the superconducting gap.\cite{Krasnov00} We
also note that the $(2 \Delta _d / k_B T_c)$ ratio in YBCO
increases with decreasing doping level, suggesting increasing
electronic correlation in the underdoped
regime.\cite{Yeh01a,Yeh01b,Yeh02}

In general, the quasiparticle spectra of the YBCO system exhibit
long-range spatial homogeneity for most doping
levels,\cite{Yeh01a,Yeh01b} in contrast to the nano-scale spectral
variations found in some underdoped BSCCO.\cite{Lang02} This
difference may be attributed to the more homogeneous oxygen
distribution in YBCO, where oxygen vacancies exist in the
CuO-chains and are often structurally ordered. In contrast, oxygen
vacancies in BSCCO can exist in different atomic planes and are
generally disordered, thereby giving rise to nano-scale
inhomogeneity in underdoped BSCCO. From the perspective of
competing orders, nano-scale phase separations can occur for
limited doping levels. Indeed, significantly more homogeneous
tunneling gap values have been reported for tunneling directly
into the CuO$_2$ plane of a BSCCO sample,\cite{Misra02} implying
that nano-scale phase separations need not be a natural
consequence of short superconducting coherence lengths.

\subsection{Effects of Quantum Impurities}
An important consequence of either $d_{x^2-y^2}$ or
$(d_{x^2-y^2}+s)$ pairing symmetry is the existence of low-energy
nodal quasiparticle excitations. These fermionic excitations can
interact strongly with quantum impurities in the CuO$_2$ plane and
significantly influence the local quasiparticle spectra near
impurities.\cite{Balatsky95,Salkola96,Flatte97,Salkola97}
Furthermore, the existence of nearest-neighbor antiferromagnetic
Cu$^{2+}$-Cu$^{2+}$ correlation in the superconducting state can
induce effective Kondo-like magnetic moments on the neighboring
Cu$^{2+}$ ions of a spinless impurity (such as Zn$^{2+}$,
Mg$^{2+}$, Al$^{3+}$ and Li$^+$ with $S=0$),\cite{Vojta01} as
confirmed from nuclear magnetic resonance
(NMR)\cite{Alloul91,Ishida96} and inelastic neutron scattering
(INS)\cite{Sidis00} experiments. These strong effects in the
p-type cuprates are in contrast to the insensitivity of
conventional superconductors to spinless
impurities\cite{Anderson59}.

\begin{figure}

\centerline{\includegraphics[width=3.5in]{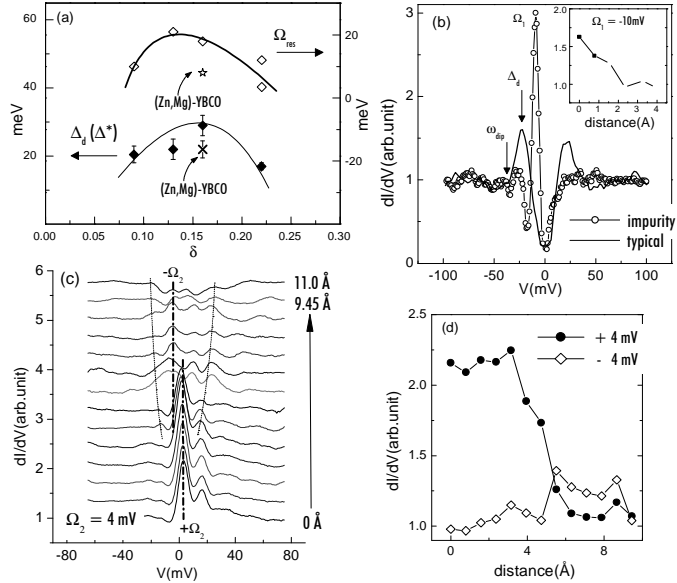}}

%

\caption{(a) Doping dependent pairing potential $\Delta _d$ and
spin excitation energy $\Omega _{res}$ of YBCO. (b) c-axis
tunneling spectra of a YBCO single crystal at and far away from a
non-magnetic impurity with a resonant scattering peak at $\Omega
_1 \approx - 10 meV$. (c) Spatial evolution of the c-axis
tunneling spectra near a non-magnetic impurity with a scattering
peak at $\Omega _2 \approx + 4 meV$. (d) Spatial evolution of the
differential conductance in (c) at $\pm \Omega _2$. All spectra
were taken at 4.2 K.} \label{fig1}
\end{figure}

Most theoretical studies of the quasiparticle tunneling spectra
near quantum impurities are restricted to perturbative and
one-band
approximation~\cite{Balatsky95,Salkola96,Flatte97,Salkola97,Vojta01}.
The Hamiltonian (${\cal H}$) is approximated by ${\cal H} = {\cal
H} _{BCS} + {\cal H} _{imp}$, where ${\cal H}_{BCS}$ is the
$d$-wave BCS Hamiltonian and ${\cal H} _{imp}$ due to impurities
contains both the potential scattering term ${\cal H} _{pot}$ and
the magnetic exchange term ${\cal H} _{mag}$:\cite{Vojta01}
\begin{equation}
{\cal H} _{imp} = {\cal H} _{pot} + {\cal H} _{mag} = U \sum _{\sigma} c_{0 \sigma} ^{\dagger} c _{0 \sigma} + \sum _{\vec R} J _{\vec R} {\bf S} \cdot {\bf \sigma _{\vec R}}.
\label{eq:H}
\end{equation}
Here $U$ is the on-site Coulomb scattering potential,
$c^{\dagger}$ and $c$ are particle operators, and $J_{\vec R}$ is
the exchange coupling constant between the spin of conduction
carriers on the $\vec R$ sites and the localized magnetic moment
($S$). If one further neglects variations in the tunneling matrix
and assumes a pure potential scattering contribution, one obtains
a resonant energy at $\Omega _0$ on the impurity site using
Eq.~(\ref{eq:H}):\cite{Balatsky95,Salkola96,Flatte97}
\begin{equation}
\vert \Omega _0 / \Delta _d \vert = (\pi /2) \cot \delta _0 \ln \left( 8 / \pi \cot \delta _0 \right),
\label{eq:pot}
\end{equation}
where $\delta _0$ is the impurity-induced phase shift in the
quasiparticle wavefunction. On the other hand, for point-like
magnetic impurities with exchange interaction limited to the
$s$-channel, one obtains two spin-polarized impurity states at
energies $\Omega _{1,2}$:\cite{Salkola97}
\begin{equation}
\vert \Omega _{1,2}/ \Delta _d \vert = 1/ \left[ 2 {\cal N} _F (U \pm W) \ln \vert 8 {\cal N} _F (U \pm W) \vert \right],
\label{eq:mag}
\end{equation}
where ${\cal N} _F$ is the density of states at the Fermi level
and $W \equiv J {\bf S} \cdot {\bf \sigma}$ assumes isolated and
equivalent magnetic impurities at all sites. In contrast, for {\it
induced} magnetic moments on the neighboring Cu$^{2+}$ sites due
to a spinless impurity, the perturbation Hamiltonian ${\cal H}
_{mag}$ must include the four Cu sites $\vec R$ adjacent to the
impurity with $J_{\vec R} = J/4$ and also involve screening
channels associated with $s$-, $p_x$-, $p_y$- and $d$-wave like
conduction electrons on the four sites.\cite{Vojta01} The
resulting quasiparticle spectra exhibit a single resonant peak at
the impurity site and alternating intensities and signs of the
peak energy away from the impurity.\cite{Vojta01}

We have performed STS studies on an optimally doped YBCO with
$0.26\%$ Zn and $0.4\%$ Mg substitution. The $T_c$ of this sample
is 82 K, substantially lower than that of optimally doped YBCO
($T_c = 93$ K). The representative spectroscopic information is
illustrated in Figs.~\ref{fig1}(b)--(d). For STM tip significantly
far away from any impurities, the tunneling spectra are similar to
typical c-axis quasiparticle tunneling spectra in pure YBCO,
although the global superconducting gap $\Delta _d$ is suppressed
to $(25 \pm 2)$ eV from the value $\Delta _d = (29 \pm 1)$ meV in
pure YBCO, as shown in Fig.~\ref{fig1}(a).\cite{Yeh01a,Yeh01b}
Moreover, the energy $\omega _{dip}$ associated with the
``dip-hump'' satellite features also shifts substantially relative
to that in pure YBCO. The dip-hump features have been attributed
to quasiparticle damping by many-body excitations such as spin
fluctuations or phonons,\cite{Chubukov00,Wu01,Lanzara01} and the
resonant energy of the many-body excitation can be empirically
determined via $|\Omega _{res}| = |\omega _{dip} - \Delta _d|$. We
find that $|\Omega _{res}|$ decreases significantly to $(7 \pm 1)$
meV from the value $(17 \pm 1)$ meV in pure YBCO. This drastic
decrease suggests that phonons are unlikely the relevant many-body
excitations that contribute to the satellite features. On the
other hand, the local spectral evolution suggests that there are
two types of impurities, one with a resonant scattering energy
$\Omega _1 \sim - 10$ meV and the other with $\Omega _2 \sim 4$
meV at the impurity site.\cite{Yeh01a,Yeh01b} The intensity of the
resonant peak decreases rapidly within approximately one Fermi
wavelength along the Cu-O bonding direction, as shown in the
insets of Figs.~\ref{fig1}(b) and (c). Moreover, the resonant
scattering peak appears to alternate between energies of the same
magnitude and opposite signs as the STM tip is displaced away from
an impurity, as exemplified in Fig.~\ref{fig1}(c). Some of these
spatially varying spectra near impurities even reveal slow
temporal variations over long times (about $\sim 10^2$ s).
Although more detailed studies and proper consideration of the
tunneling matrix\cite{Zhu00,Vojta01} are needed to fully
understand the relative contributions of ${\cal H} _{pot}$ and
${\cal H} _{mag}$, the presence of temporal variations might be
more consistent with the Kondo effect. For comparison with other
systems, however, we may assume pure potential scattering and use
Eq.~(\ref{eq:pot}) to derive the corresponding phase shifts
associated with the spinless quantum impurities. We obtain $\delta
_1 \sim 0.38 \pi$ and $\delta _2 \sim 0.43 \pi$ in comparison with
the phase shift ($\delta _0 \sim 0.45 \pi$) due to Zn in
BSCCO\cite{Pan00}, suggesting that the impurity scattering
strength in YBCO is weaker.

\section{PAIRING STATE OF THE N-TYPE CUPRATES}
Despite strong consensus for the pairing symmetry in p-type
cuprates as being predominantly
$d_{x^2-y^2}$,\cite{vanHarlingen95,Tsuei00a} the situation
associated with the n-type cuprates remains controversial. In the
case of one-layer n-type cuprates $\rm Ln_{2-x}M_xCuO_4$ system
(Ln = Nd, Sm, Pr; M = Ce, Sr), tunneling spectroscopic studies on
bulk materials are consistent with $s$-wave pairing\cite{Alff99}
whereas phase sensitive measurements on thin films suggest
$d_{x^2-y^2}$-pairing.\cite{Tsuei00b}. Recently, further studies
of the one-layer n-type cuprates have implied doping dependent
pairing symmetry.\cite{Skinta02}

\subsection{Strongly Correlated S-Wave Pairing}
\begin{figure}
\centerline{\includegraphics[width=3in, height=3in]{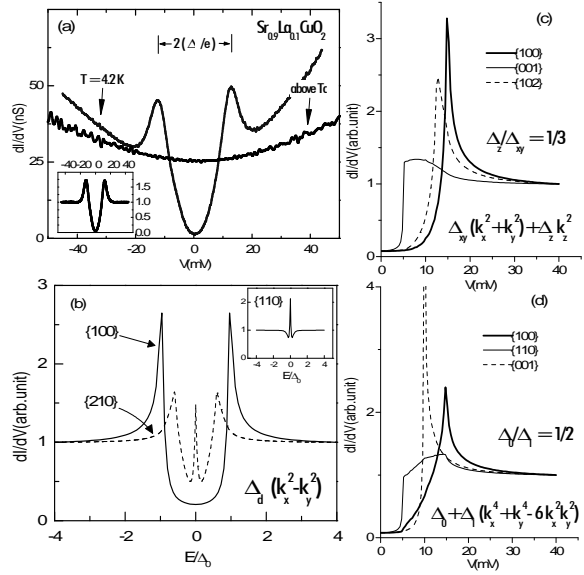}}
%
%
\caption{(a) Normalized momentum-independent quasiparticle
tunneling spectra in a pure $\rm Sr_{0.9}La_{0.1}CuO_2$ taken at
$T = 4.2$ K. (b) Calculated spectra for quasiparticle tunneling
along different crystalline axes of a pure $d_{x^2-y^2}$-wave
superconductor. (c) Calculated spectra for an anisotropic $s$-wave
pairing potential with uniaxial symmetry. (d) Calculated spectra
for an anisotropic $s$-wave pairing potential with 4-fold in-plane
modulation.}

\label{fig2}
\end{figure}

Our STS studies of the simplest form of cuprate superconductors,
the n-type ``infinite-layer'' system $\rm Sr_{0.9}La_{0.1}CuO_2$
(SLCO) with $T_c$ = 43 K\cite{Jung02a} reveal momentum-independent
quasiparticle tunneling spectra, as manifested by consistent
spectral characteristics among data taken on more than 300
randomly oriented grains of a polycrystalline SLCO
sample.\cite{Yeh02,Chen02} A representative spectrum is shown in
Fig.~\ref{fig2}(a). This finding together with the absence of any
$d$-wave spectral characteristics (see Fig.~\ref{fig2}(b) with
$\Delta _k = \Delta _d (k_x ^2 - k_y ^2)$) is suggestive of
$s$-wave pairing symmetry.\cite{Chen02} To further investigate the
possibility of any anisotropy, we use the generalized BTK analysis
to derive tunneling spectra for the anisotropic pairing potentials
permitted by symmetry. For the $D_{4h}$ point group associated
with the infinite-layer system, we may consider the lowest energy
configurations in the pairing state with angular momenta $\ell =
0, 2, 4$. We find that the pairing potential of the $A_{1g}$
representation and $\ell = 0$ corresponds to the isotropic
$s$-wave pairing potential $\Delta _0$, and that $\ell$ up to 2 is
given by $\Delta _k = \Delta _{xy} (k_x ^2 + k_y ^2) + \Delta _z
k_z^2$, where $\Delta _{xy}$ and $\Delta _z$ are the in-plane and
c-axis pairing potentials. For $\ell$ up to 4, two pairing
potentials of the $A _{1g}$ representation are possible. One is
similar to that for $\ell = 2$ with uniaxial symmetry, the other
exhibits four-fold modulations in the $k_x$-$k_y$ plane, with
$\Delta _k = \Delta _0 + \Delta _1 (k_x ^4 + k_y ^4 - 6 k_x ^2 k_y
^2)$. The tunneling spectra for these possibilities are shown in
Figs.~\ref{fig2}(c) and (d), which suggest that all pairing
potentials with anisotropy would have resulted in significantly
varying tunneling gap values with the quasiparticle momenta.
Comparing this analysis with our data and experimental resolution,
we estimate $\sim 8 \%$ upper bound for any anisotropy in the
pairing potential.

The finding of $s$-wave pairing symmetry may be related to the
fact that the n-type infinite-layer system is the only cuprate
with a c-axis superconducting coherence length longer than the
c-axis lattice constant.\cite{Chen02,Jung02a,Jung02b} This unique
property is in contrast to the quasi-two dimensional
superconductivity in all other cuprates. Additional noteworthy
phenomena include insignificant satellite features, which imply
much reduced spin fluctuations below $T_c$, and the absence of
pseudogap above $T_c$, as shown in Fig.~\ref{fig2}(a). The
insignificant spin fluctuations are consistent with the fact that
electron doping results in formation of $Cu^+$-ions that dilute
the antiferromagnetic Cu$^{2+}$-Cu$^{2+}$ correlation without
causing significant spin fluctuations as holes do in the oxygen
$p$-orbitals, and the absence of pseudogap above $T_c$ is also
consistent with similar findings in the one-layer n-type
cuprates.\cite{Alff02}

\begin{figure}
\centerline{\includegraphics[width=3.25in]{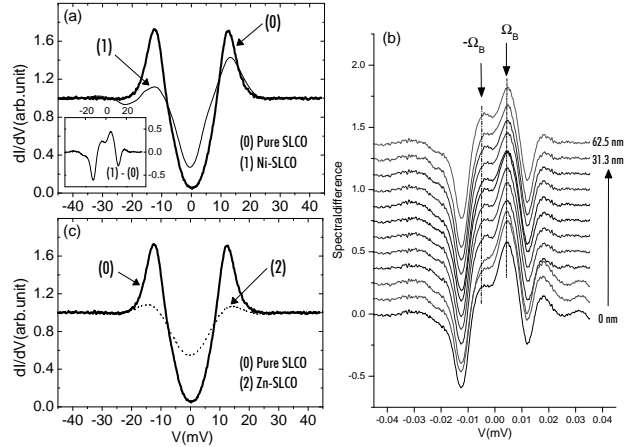}}
%
%

\caption{(a) Comparison of the quasiparticle tunneling spectra of
$\rm Sr_{0.9}La_{0.1}CuO_2$ and $\rm
Sr_{0.9}La_{0.1}(Cu_{0.99}Ni_{0.01})O_2$ at $T = 4.2$ K. (b)
Similar comparison for $\rm Sr_{0.9}La_{0.1}CuO_2$ and $\rm
Sr_{0.9}La_{0.1}(Cu_{0.99}Zn_{0.01})O_2$. (c) Spectral difference
due to Ni-impurities, showing long-range impurity bound states at
$\pm \Omega _B$, similar to the Shiba impurity bands.}

\label{fig3}
\end{figure}

\subsection{Effects of Quantum Impurities}
In contrast to the sensitive response of p-type cuprates to both
magnetic and non-magnetic impurities, the response of the
infinite-layer system appear to be more consistent with that in
conventional superconductors.\cite{Anderson59,Abrikosov61,Shiba68}
That is, little suppression in either $T_c$ or the superconducting
gap is found with Zn$^{2+}$-impurity substitution up to
3\%.\cite{Jung02b,Chen02} In contrast, strong $T_c$ suppression
and significant impurity-induced electron-hole asymmetry in the
quasiparticle spectra were found with small concentrations of
Ni$^{2+}$-impurities.\cite{Jung02b,Chen02} Moreover, our studies
of the local quasiparticle density of states in the 1\%
Ni-substituted SLCO sample reveal long-range impurity effects,
with strong particle-hole asymmetry due to the magnetic
impurity-induced broken time-reversal symmetry, as shown in
Fig.~\ref{fig3}(a). In contrast, 1 \% Zn-isubsitutions result in
significant disorder states without reduction in either the
superconducting gap (Fig.~\ref{fig3}(b)) or $T_c$.

The inset of Fig.~\ref{fig3}(a) illustrates the spectral
contribution due to Ni-impurities, and the corresponding gradual
spatial evolution is shown in Fig.~\ref{fig3}(c). Such impurity
spectra are in contrast to the rapid spatial variations in
(Zn,Mg)-substituted YBCO (see Fig.~\ref{fig1}(c))\cite{Yeh01a} and
in Ni-substituted BSCCO.\cite{Hudson01} The spectral contributions
of Ni-impurities resemble the Shiba states for magnetic impurity
bands in $s$-wave superconductors,\cite{Shiba68} and the bound
state energies are found to peak at $\pm \Omega _B$, where
\begin{equation}
\vert \Omega _B / \Delta _0 \vert = (\pi /2)JS {\cal N}_F \equiv \zeta. \label{eq:Shiba}
\end{equation}
Equation~(\ref{eq:Shiba}) clearly differs from Eq.~(\ref{eq:mag}),
the latter being associated with magnetic impurities in $d$-wave
superconductors. Using $\Omega _B \sim 5$ meV and $\Delta _0 \sim
13$ meV, we obtain $\zeta \sim 2/3$. The assumption of impurity
bands can be justified by noting that the average Ni-Ni separation
($\sim 1.8$ nm) is shorter than the in-plane superconducting
coherence lengths of SLCO ($\sim 4.8$ nm),\cite{Jung02a} such that
substantial overlap of impurity wavefunctions can be expected.

\section{SUMMARY}
Our STS studies of both p-type and n-type cuprates reveal that the
pairing symmetry, pseudogap phenomenon and spin fluctuations are
in fact not universal. The only ubiquitous features among all
cuprates appear to be the strong electronic correlation and
nearest-neighbor Cu$^{2+}$-Cu$^{2+}$ antiferromagnetic interaction
in the CuO$_2$ planes.

\section*{ACKNOWLEDGMENTS}
This research is jointly supported by NSF grant DMR-0103045 at
Caltech, by NASA at the Jet Propulsion Laboratory, by the Ministry
of Science and Technology of Korea at the Pohang University, and
by NEDO at SRL/ISTEC in Japan.

\end{document}